\documentclass[12pt]{iopart}

\usepackage[colorlinks=true,linkcolor=blue,urlcolor=blue,citecolor=blue]{hyperref}
\usepackage{graphicx}

\begin{document}

\title[]{Hyperfine Paschen-Back regime of Potassium D$_2$ line observed by Doppler-free spectroscopy}

\author{A. Sargsyan$^1$, E. Klinger$^{1,2}$, A. Tonoyan$^1$, C. Leroy$^2$, \& D. Sarkisyan$^1$}

\address{$^1$ Institute for Physical Research, NAS of Armenia, 0203 Ashtarak-2, Armenia}
\address{$^2$ Laboratoire Interdisciplinaire Carnot de Bourgogne, UMR CNRS 6303, Universit\'e Bourgogne Franche-Comt\'e, Dijon, France}
\ead{emmanuel.klinger@u-bourgogne.fr}
\vspace{10pt}
\begin{indented}
\item[]
\end{indented}

\begin{abstract}
Selective reflection of a laser radiation from an interface formed by a dielectric window and a potassium atomic vapour confined in a nano-cell with 350~nm gap thickness is implemented for the first time to study the atomic transitions of K D$_2$ line in external magnetic fields. In moderate $B$-fields, there are 44 individual Zeeman transitions which reduce to two groups (one formed by $\sigma^+$ the other one by $\sigma^-$ circularly-polarised light), each containing eight atomic transitions, as the magnetic field increases. Each of these groups contains one so-called ``guiding'' transition whose particularities are to have a probability (intensity) as well as a frequency shift slope (in MHz/G) that are constant in the whole range of 0 -- 10~kG  magnetic fields. In the case of $\pi$-polarised laser radiation, among eight transitions two are forbidden at $B = 0$, yet their probabilities undergo a giant modification under the influence of a magnetic field. We demonstrate that for $B$-fields $> 165$~G a complete hyperfine Paschen-Back regime is observed. Other peculiarities of K D$_2$ line behaviour in magnetic field are also presented. We show a very good agreement between theoretical calculations and experiments. The recording of the hyperfine Paschen-Back regime of K D$_2$ line with high spectral resolution is demonstrated for the first time. 
\end{abstract}

%

%
%
%

\section{Introduction}
It has recently been demonstrated that the selective reflection (SR) of a laser radiation from an interface of dielectric window and atomic vapour confined in a nano-cell (NC) with a thickness of a few hundred nanometres is a convenient tool for atomic spectroscopy \cite{sargsyan_jetpl_2016,sargsyan_josab_2017,sargsyan_ol_2017}. The real-time derivative of SR signal (dSR), where each frequency position of the recorded peaks coincides with the atomic transition ones, is used and provides a 30 -- 50~MHz spectral resolution with linearity of the signal response with respect to the transition probabilities. The large amplitude and the sub-Doppler width of a detected signal in addition to the simplicity of the dSR-method make it appropriate for applications in metrology and magnetometry. In particular, the dSR-method provides a convenient frequency marker for atomic transitions \cite{sargsyan_jetpl_2016}. In \cite{sargsyan_ol_2017}, we have implemented the dSR-method for atomic layers having thicknesses of a few tens of nanometres to probe atom-surface interactions and we have observed a 240 MHz red-shift for a cell thickness $L\sim40$~nm. With the dSR-method, a complete frequency-resolved hyperfine Paschen-Back (HPB) splitting of ten atomic transitions (four for $^{87}$Rb and six for $^{85}$Rb) was recorded in a strong magnetic field ($B > 2$~kG) in \cite{sargsyan_jetpl_2016}; similar results for Cs have been reported in \cite{sargsyan_josab_2017}. 

One of the reason why K atomic vapours are less frequently used as compared to Rb or Cs ones is the following: for a temperature $\sim100~^\circ$C the Doppler-broadening is $\sim0.9$~GHz, which exceeds the hyperfine splitting of the ground and excited levels ($\sim462$~ MHz and $\sim10-20$~MHz, correspondingly) such that the transitions $F_g=1, 2\rightarrow F_e= 0,1,2,3$ of $^{39}$K are completely masked by the Doppler-profile. Therefore, there is a small number of papers concerning the laser spectroscopy of potassium: the accurate identification of atomic transitions of K was reported in \cite{das_jpb_2008,hanley_jpb_2015}; saturated absorption spectra of the D$_1$ line of potassium atoms have been studied in details both theoretically and experimentally in \cite{bloch_lp_1996}. Potassium vapours were used for the investigation of nonlinear magneto-optical Faraday rotation in a antirelaxation paraffin-coated cell \cite{guzman_pra_2006}, polarisation spectroscopy and magnetically-induced dichroism for magnetic fields in the range of 1 -- 50~G \cite{pahwa_oe_2012}, formation of dark resonance having a sub-natural linewidth \cite{lampis_oe_2016}, electromagnetically induced transparency \cite{sargsyan_os_2017} and four-wave mixing process \cite{zlatkovic_lpl_2016}. A theory describing the transmission of Faraday filters based on sodium and potassium vapours is presented in \cite{harrell_josab_2009}.

In external magnetic field, there is an additional splitting of the energy levels which causes the formation of a number of atomic transitions spaced by a frequency interval of $\sim100$~MHz in the HPB regime; that is why a Doppler-free method must be implemented to perform an efficient study of atomic transitions of K vapours. In this paper we demonstrate that the dSR-method is a very convenient one to investigate the behaviour of an individual transition of $^{39}$K D$_2$ line (energy levels are shown in Fig~\ref{fig1}a). The particularity of $^{39}$K  is a small characteristic value of magnetic field $B_0 =A_{hf}/\mu_B =165$~G at which the HPB regime starts as compared to other alkalis such as  $^{133}$Cs ($B_0 =1700$~G) and $^{ 85}$Rb ($B_0 =700$~G), $^{87}$Rb ($B_0 =2400$~G) isotopes \cite{olsen_pra_2011,sargsyan_ol_2012,weller_ol_2012,zentile_cpc_2015}; where $A_{hf}$ is the magnetic dipole constant for $4^2S_{1/2}$ ground level and $\mu_B$ is the Bohr magneton. Hence, there is a significant change in atomic transition probabilities of $^{39}$K D$_2$ line for relatively small magnetic fields, at least by an order smaller as compared to other alkalis. To our best knowledge, the recording of the HPB regime of K D$_2$ line with high spectral resolution is demonstrated for the first time.

\section{Experimental arrangement}

\begin{figure}[ht]
\centering
\includegraphics[width=0.9\textwidth]{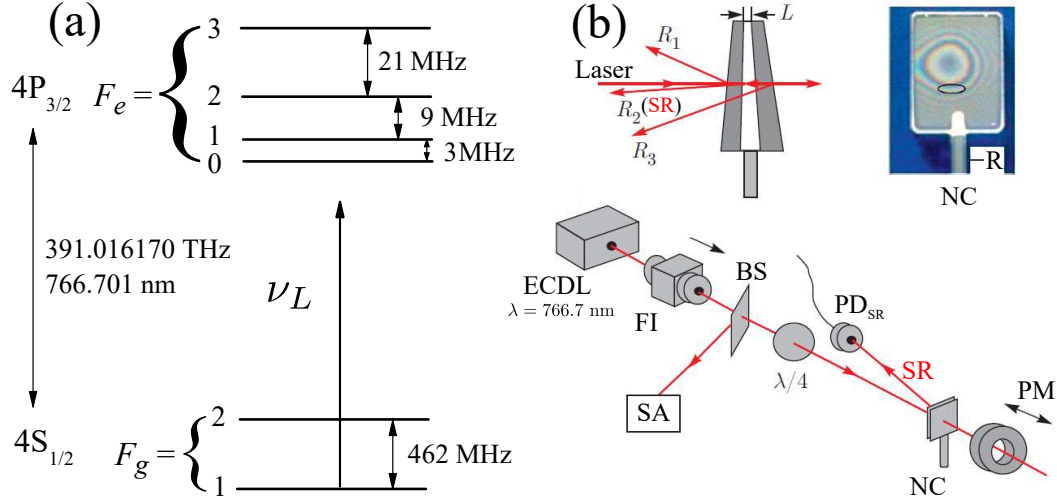}
\caption{(a) Energy-level diagram for $4^2S_{1/2}$ and $4^2P_{3/2}$ states of  $^{39}$K. (b) Layout of the experimental setup: ECDL -- extended cavity diode laser, FI -- Faraday isolator, BS -- beam splitter, NC -- K nano-cell inside an oven (not shown), (R) -- thin sapphire reservoir, PM -- permanent magnet, PD -- photodetector, $\lambda /4$ -- quarter-wave plate, SR -- selective reflection channel, SA -- saturated absorption reference channel. Upper right inset: photograph of the NC; the oval marks the region $L=250 - 350$~nm. Upper left inset: geometry of the 3 reflected beams; the selective reflection beam (SR) propagates in the direction of $R_2$.}
\label{fig1}
\end{figure}

Figure \ref{fig1}b shows the layout of the experimental setup.  A frequency-tunable cw narrowband  ($\gamma_L \sim 2\pi \cdot 1$~MHz) extended cavity diode laser (ECDL) with $\lambda =766.7$~nm wavelength, protected by a Faraday isolator (FI),  emits a linearly polarised radiation directed at normal incidence onto a K nano-cell mounted inside the oven. A quarter-wave plate, placed in between the FI and the NC, allows to switch between $\sigma^+$ (left-hand) and $\sigma^-$ (right-hand) circularly-polarised radiation. The NC is filled with natural potassium that consist of $^{39}$K (93.25\%) and $^{41}$K (6.70\%) atoms, and the details of its design can be found in \cite{keaveney_prl_2012,sargsyan_epl_2015}. The necessary vapour density $N\sim 5\times 10^{12}$~cm$^{-3}$ was attained by heating the cell's thin sapphire reservoir (R) containing metallic potassium, to $T_R \sim 150~^\circ$C, while keeping the window temperature some $20~^\circ$C higher.

A longitudinal magnetic field $\mathbf{B}\parallel\mathbf{k}$ up to 1~kG, where $\mathbf{k}$ is the wavevector of the laser radiation, was applied using a permanent neodymium-iron-boron alloy magnet placed near the output window of the NC. The variation of the field strength was achieved by axial displacement of the magnet system and was monitored by a calibrated magnetometer. In spite of a strong spatial gradient of the field produced by the permanent magnet, the field inside the interaction region is uniform thanks to the very small thickness of the NC. The right inset of Fig.~\ref{fig1}b shows the photograph of the K NC where one can see interference fringes formed by light reflection from the inner surfaces of the windows because of variable thickness $L$ of the vapour column across the aperture.

The SR measurements (geometry of the reflected laser beams presented in the left inset) were performed for $L \sim 350$~nm. Although the decrease of $L$ improves the spatial resolution (which is very important when using high-gradient field permanent magnets), it simultaneously results in a broadening of the SR spectral linewidth; therefore $L = 350$~nm appears as the optimal thickness. This additional broadening is a result of atom-wall collisions: the reduction of the thickness $L$ between the windows shortens the flight time of atoms toward the surface, determined as $t = L/v_z $ ($v_z$ is the projection of the thermal velocity perpendicular to the window plane), thus making atom-wall collisions more frequent and leading to additional broadening. To form a frequency reference, a part of the laser radiation was branched to an auxiliary saturated absorption (SA) setup formed in a 1.4~cm-long K cell.

\section{Theoretical considerations}

In this section, we give the outline of our theoretical model, additional details on the calculations of dSR spectrum are given in \cite{sargsyan_josab_2017,klinger_epjd_2017}. The problem of calculating the spectrum of alkaline vapours confined in a NC when a longitudinal $B$-field is applied can be split in two points: the calculation of the transition probabilities and frequencies \cite{tremblay_pra_1990}, and the calculation of the line profile inherent to the NC properties \cite{zambon_oc_1997,dutier_josab_2003}.

\subsection{Transition probabilities and frequencies under magnetic field}

The starting point for spectroscopic analysis of alkali vapours under longitudinal magnetic field is to write down the Hamiltonian of the system $H_m$ as the sum of the hyperfine structure Hamiltonian $H_0$ modified by the magnetic interaction, such that
\begin{equation}
H_m = H_0 +\frac{\mu_B}{\hbar}B_z(g_LL_z + g_SS_z+g_II_z),
\label{eq:hamil_B}
\end{equation} 
where $L_z$, $S_z$, and $I_z$ are respectively the projection of the orbital, electron spin and nucleus spin momentum along $z$, chosen as the quantization axis; $g_{L,S,I}$ are the associated Land\'e factors (for the sign convention, see \cite{steck_2011}). Details of the construction of the Hamiltonian in the base $|F,m_F\rangle$ can be found in \cite{tremblay_pra_1990}. The transitions probabilities $W_{eg}$ are proportional to the square of the dipole moment $\mu_{eg}$ between the states $|e\rangle$ and $|g\rangle$
\begin{equation}
W_{eg}\propto \left(\sum_{F_e'F_g'}c_{F_eF_e'}a(\Psi(F_e,m_e);\Psi(F_g,m_g);q)c_{F_gF_g'}\right)^2,
\end{equation}
with the coefficients $c_{FF'}$ given by the eigenvectors of diagonalized $H_m$ matrix, and
\begin{eqnarray}
\fl \eqalign{a(\Psi(F_e,m_e);\Psi(F_g,m_g);q)=&(-1)^{1+I+J_e+F_e+F_g-m_{Fe}}\sqrt{2J_e+1}\sqrt{2F_e+1}\sqrt{2F_g+1}\\
&\times \left( \begin{array}{r@{\quad}cr} 
F_e & 1 & F_g \\
-m_{F_e} & q & m_{F_g}
\end{array}\right)
\left\{ \begin{array}{r@{\quad}cr} 
F_e & 1 & F_g \\
J_g & I & J_e
\end{array}\right\},}
\end{eqnarray}
where the parentheses and the curly brackets denote the 3-$j$ and 6-$j$ coefficients, respectively; $q=0,\pm1$ is associated the polarisation of the excitation such that $q=0$ for a $\pi$-polarised laser field, $q=\pm1$ for a $\sigma^\pm$-polarised laser field. 

\subsection{Line profile}

The line profile of atoms confined in thin cells having a gap between their windows of the order of the laser wavelength has been deeply studied in \cite{zambon_oc_1997,dutier_josab_2003}. The resonant contribution to the reflected field reads (in intensity)
\begin{equation}
S_r\cong 2 \frac{t_{cw}}{|Q|^2}\Re{\Big\lbrace r_w\big[1-\exp(-2ikL)\big]\times\big[I_b - r_wI_f\exp(2ikL)\big]\Big\rbrace}E_{in},
\end{equation}
where $t_{cw}$, $r_w$ are  respectively transmission and reflection coefficients, $Q=1-r_w^2\exp(2ikL)$ is the quality factor associated to the NC of thickness $L$; $I_f$ and $I_b$ are integrates of the forward and backward polarisation
\begin{equation}
I_f=\frac{ik}{2\epsilon_0}\int_0^L P_0(z)dz, \qquad I_b=\frac{ik}{2\epsilon_0}\int_0^L P_0(z)\exp(2ikz)dz.
\end{equation}
The polarisation $P_0(z)$ is induced by the interaction of the laser with an ensemble of 2-level systems; it is given by averaging the coherences of the reduced density matrix $\rho$ over the atomic distribution of speed in the cell (supposed as Maxwellian)
\begin{equation}
P_{0}(z) = \sum_i N\mu_{i}\int\limits^{+\infty}_{-\infty} W(v)\rho^i_{eg}(z,v,\Delta_i) dv,
\label{eq:polarization_only_positive_velocities_Ensemble_2-level_system}
\end{equation}
with $\Delta_i=\omega-\omega_i$. Each 2-level system $|i\rangle$, having a transition frequency $\nu_i=\omega_i/2\pi$ and a transition intensity $|\mu_i|^2$, can contribute to the recorded signal if they are close to resonance with the laser pulsation $\omega$. Expressions of the atomic coherences $\rho^i_{eg}$ are found by solving the Liouville equation of motion of the density matrix, that is
\begin{equation}
\frac{d}{dt}\rho=-\frac{i}{\hbar}\big[H,\rho\big]-\frac{1}{2}\big\lbrace \Gamma,\rho\big\rbrace,
\end{equation}
where $H$ is the Hamiltonian describing the interaction of the vapour with the laser radiation, and the matrix $\Gamma$ accounts for homogeneous relaxation processes; $\lbrace a, b \rbrace= ab + ba$ is the anticommutator.

\section{Results and discussion}
\subsection{Circular polarisation analysis}
\begin{figure}[ht]
\centering
\includegraphics[width=0.99\textwidth]{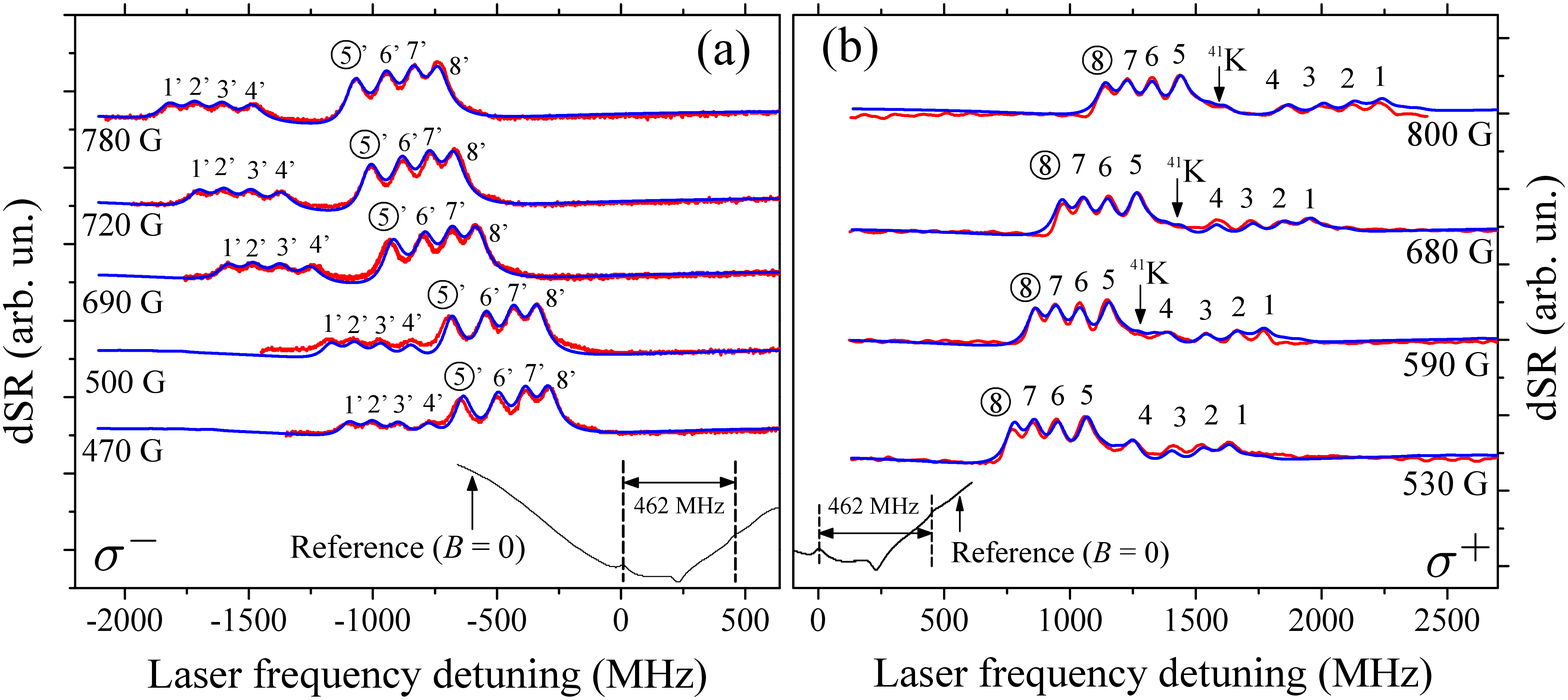}
\caption{$^{39}$K D$_2$ line recorded (red solid lines) and calculated (blue solid lines) spectra for (a) $\sigma^-$-polarised light at $B = 470,~500,~690,~720,~ 780$~G; and (b) $\sigma^+$-polarised laser radiation at $B = 530,~590,~680,~800$~G. Experimental parameters: transition linewidth $\sim 100$~MHz, NC's thickness $L=350$~nm, laser power $P_L=0.1$~mW, reservoir's temperature $T_R = 150~^\circ$C. The lower curve is the recorded SA spectrum that serves as a frequency reference. The curves have been shifted vertically for clarity. In each case, all eight atomic transitions are well spectrally resolved. Although the proportion of $^{41}$K in the cell is small (6.70\%), a portion of its spectra can be seen; it is indicated on the graphs with arrows.}
\label{fig2}
\end{figure}

On Fig.~\ref{fig2}a, the red curves show dSR experimental spectra in the case of $\sigma^-$ circularly-polarised laser radiation for five different values of the applied longitudinal magnetic field, from bottom to top: 470, 500, 690, 720 and 780~G. The complementary study with a $\sigma^+$  circularly-polarised excitation for the $B$-field values of 530, 590, 680 and 800~G is shown on Fig.~\ref{fig2}b. The spectra are recorded for a reservoir temperature $T_R\sim150~^\circ$C, a laser power $P_L\sim 0.1$~mW, and atomic transitions linewidth $\sim80$~MHz Full Width at Half Maximum (FWHM).

\begin{figure}[ht]
\centering
\includegraphics[width=0.99\textwidth]{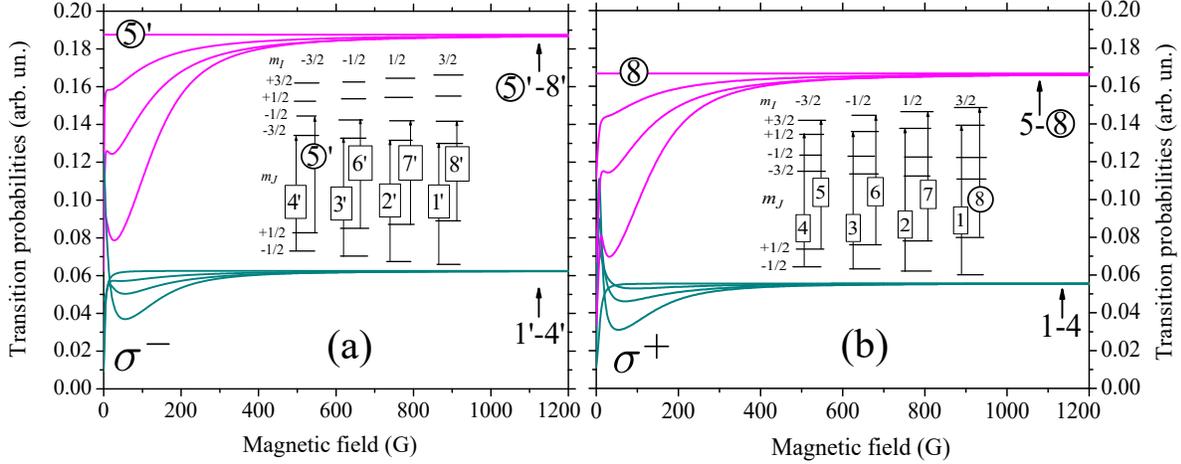}
\caption{(a) Evolution of the probabilities of 1'--4' and \textcircled{5}'--8' transitions versus $B$-field for a $\sigma^-$-polarised excitation. (b) Evolution of the probabilities of 1--4 and 5--\textcircled{8} transitions versus $B$-field for a $\sigma^+$-polarised excitation. The insets show the corresponding atomic transitions diagrams of $^{39}$K D$_2$ line in the HPB regime, expressed in the basis $|m_J, m_I\rangle$ (uncoupled basis). Selection rules for the transitions are $\Delta m_J =\pm1,~ \Delta m_I = 0$ for $\sigma^\pm$-polarised light. For simplicity, only the transitions that remain in the spectrum for strong magnetic field are presented.}
\label{fig3}
\end{figure}

It is worth to note that the dSR amplitudes are proportional to the relative probabilities presented in Fig.~\ref{fig3}. As it is seen from Fig.~\ref{fig2}a, there are two groups formed by transitions 1'--4' and \textcircled{5}'--8' and all these eight transitions whose diagram is shown in the inset of Fig.~\ref{fig3}a are well seen. The same remark holds for transitions 1--\textcircled{8} from Fig.~\ref{fig2}b (transition diagram shown in the inset of Fig.~\ref{fig3}b). Note that for a given group, the amplitudes of the transitions are equal to each other with the frequency intervals between them beeing nearly equidistant. These peculiarities as well as the strict number of atomic transitions which remain the same with increasing magnetic field are evidences of the establishment of Pashen-Back regime. Transitions labeled  \textcircled{5}' and \textcircled{8} are the so-called ``guiding'' transitions (GT) \cite{sargsyan_epl_2015,sargsyan_jetpl_2015}: their probability as well as their frequency shift slope remain the same ($s^\pm=\pm1.4$~MHz/G) in the whole range of applied $B$-fields.

The lower (black) curves show SA spectra for $B=0$. As shown in \cite{zielinska_ol_2012}, the existence of crossover lines makes the SA technique useless for a spectroscopic analysis for $B>100$~G. The blue curves show the calculated dSR spectra of $^{39}$K and $^{41}$K isotopes with the linewidth of 80 -- 100~MHz for $\sigma^-$- (Fig.~\ref{fig2}a) and $\sigma^+$- (Fig.~\ref{fig2}b) polarised laser radiations. As it is seen, there is a very good agreement between the experiment and the theory. Although there is only 6.70\% of $^{41}$K isotope in natural K, a much better agreement with the experiment is realized when $^{41}$K levels are also included into theoretical considerations; particularly the peaks shown by the arrows in Fig.~\ref{fig2}b are caused by the influence of $^{41}$K isotope. A very good agreement between the experiment and the theory can be seen for both polarisations and all applied magnetic fields. 

It is important to note that at a relatively small magnetic field $B\sim 400$~G the two groups are already well formed and separated which is, as mention earlier, a consequence of the small value of $B_0(^{39}$K$)=165$~G. In order to detect similar well formed groups for $^{87}$Rb atoms, one must apply a much stronger magnetic field of $B\sim 6$~kG since  $B_0(^{87}$Rb$)/ B_0(^{39}$K$) \sim 15$. It is also interesting to note that the total number of the atomic transitions for both circularly-polarised laser excitations is 44 when $B\sim B_0(^{39}$K$) \sim 150$~G, while for $B\gg B_0(^{39}$K) only 16 transitions remain: this is the manifestation of the HPB regime. \\

Figure \ref{fig3} shows the transition probabilities versus magnetic field for (a) 1'--4' and \textcircled{5}'--8' transitions ($\sigma^-$ excitation)  and  for (b)  1--4  and 5--\textcircled{8} transitions ($\sigma^+$ excitation); for labelling, see the transition diagrams in insets. As we see, the transition probabilities inside the groups 1--4 and 1'--4' for $B\gg 165$~G tend asymptotically to the same value (respectively 0.056  and 0.0625); the same remark can be made for the groups 5--\textcircled{8} and \textcircled{5}'--8' which probabilities tend to the ones of the GT \textcircled{8} (0.167) and \textcircled{5}' (0.185) respectively. The later remarks are another peculiarities of the HPB regime.

\begin{figure}[ht]
\centering
\includegraphics[width=0.7\textwidth]{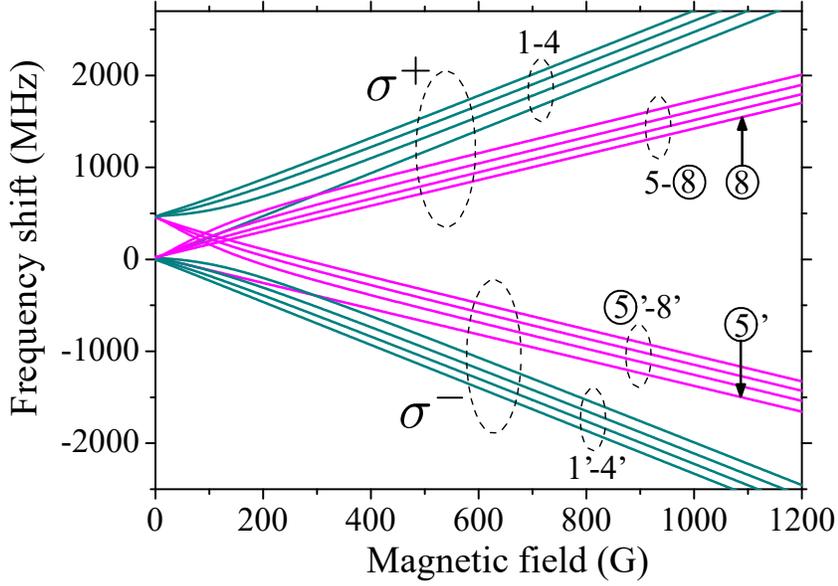}
\caption{Calculated magnetic field dependence of the frequency shifts of $^{39}$K D$_2$ line, for the transitions 1--4 and 5--\textcircled{8} ($\sigma^+$ excitation) and  for the transitions 1'--4'  and \textcircled{5}'--8' ($\sigma^-$ excitation). The guiding transitions \textcircled{8} and \textcircled{5}' are indicated.}
\label{fig4}
\end{figure}
The calculated magnetic field dependence of the transition frequency shifts under circularly-polarised excitation is presented on Fig.~\ref{fig4}. Note that the frequency slope of transitions 5--\textcircled{8} tends asymptotically to the same value of the GT transition \textcircled{8} ($s^+=+1.4$~MHz/G), while the frequency slope of transitions  \textcircled{5}'  --8′ transitions tends to the one of the GT transition \textcircled{5}'   ($s^-=-1.4$~MHz/G).

\subsection{Linear polarisation analysis}

To achieve a linear ($\pi$) polarisation excitation of the K vapour, the experimental setup presented on Fig.~\ref{fig1}b is slightly modified: the $B$-field is directed along the laser electric field $\mathbf{E}$ ($\mathbf{B} \perp \mathbf{k}$), the $\lambda /4$ plate is removed and two permanent magnets are used to set the $\mathbf{B} \parallel \mathbf{E}$ configuration. 

\begin{figure}[ht]
\centering
\includegraphics[width=0.7\textwidth]{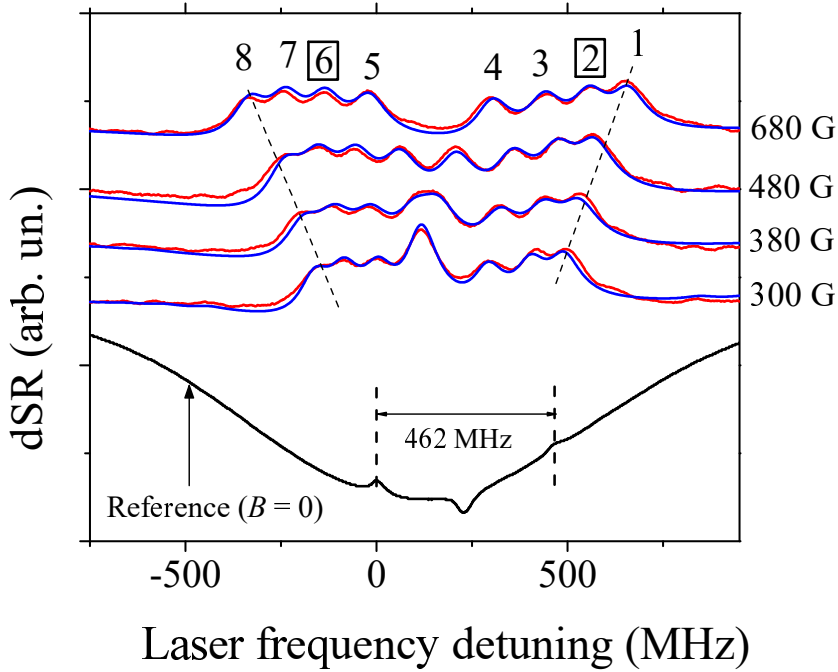}
\caption{K D$_2$ line, for $\pi$-polarised radiation recorded and calculated for $B=300,~380,~480,~680$~G. The red and blue traces show respectively the experimental and theoretical dSR spectra of $^{39}$K and $^{41}$K atoms, with a linewidth $\sim120$~MHz, NC's thickness $L=350$~nm, $P_L=0.1$~mW, and $T_R \sim 150~^\circ$C. The lower curve is the SA spectrum that serves as a frequency reference. The transitions $\fbox{2}$ and $\fbox{6}$ are IFFA transitions (see the text). The dashed lines show the frequency position of the atomic transitions and are drawn for convenience.}
\label{fig5}
\end{figure}
\begin{figure}[ht]
\centering
\includegraphics[width=0.99\textwidth]{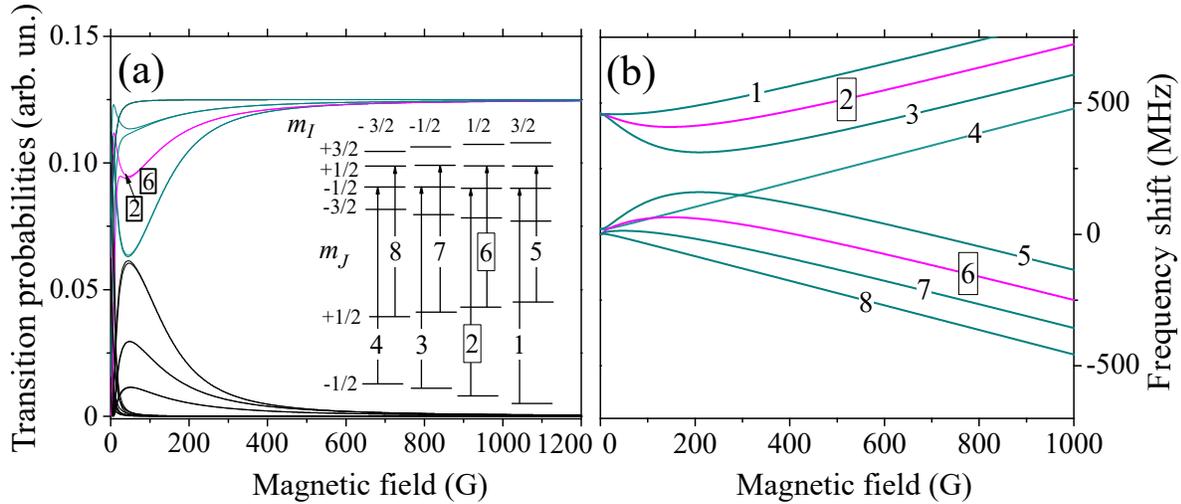}
\caption{Calculated probabilities (a) and frequency shifts (b) of $^{39}$K D$_2$ line  $F_g =1, 2 \rightarrow F_e=0,1,2,3$ transitions for $\pi$-polarised laser radiation. Transition diagram (in the uncoupled basis $|m_I,m_J\rangle$) for the HPB regime is shown in the inset. Selection rules for the transitions are $\Delta m_J =0,~ \Delta m_I = 0$. The transitions $\fbox{2}$ and $\fbox{6}$ are IFFA transitions (see the text).}
\label{fig6}
\end{figure}

In Fig.~\ref{fig5}, the red curves represent dSR experimental spectra obtained for $B=300,~380,~480,~680$~G, with $T_R\sim150~^\circ$C and a $\pi$-polarised laser radiation having a power $P_L=0.1$~mW. The blue lines show the theoretical calculations with the corresponding experimental parameters. As we see there are eight well resolved transitions, labelled 1--8, having amplitudes that tend asymptotically to the same value (see Fig.~\ref{fig6}a). Note that, in this case, the transition linewidth is a bit larger ($\sim120~$MHz) which is caused by inhomogeneities of the transverse magnetic field across the laser beam diameter of 1~mm. The magnetic field dependence of their transition frequencies is presented on Fig.~\ref{fig6}b.

The transitions $\fbox{2}$ and $\fbox{6}$ are $|F_g=1, m_F=0\rangle\rightarrow |F_e=1, m_F=0\rangle$ and $|F_g=2, m_F=0\rangle\rightarrow |F_e=2, m_F=0\rangle$ transitions. For zero magnetic field dipole matrix elements for these  $\pi$ transitions are zero \cite{steck_2011}, in other words they are ``forbidden'': for these transitions neither resonant absorption nor resonant fluorescence is detectable. In Fig.~\ref{fig6}a, the transition probability of $\fbox{2}$ and $\fbox{6}$ starts from zero and undergo a significant enhancement with increasing magnetic field. For this reason, we call them``initially forbidden further allowed'' (IFFA) transitions. Note that there is another type of ``forbidden'' transitions, so-called magnetically induced (MI) transitions since they appear only with the presence of $B$-field and verify the selection rule $\Delta F = F_e-F_g= \pm2$; particularly, the groups of transitions $F_g=2\rightarrow F_e=4$ (Rb D$_2$ line) and $F_g=3\rightarrow F_e=5$ (Cs D$_2$ line) have been respectively studied in \cite{klinger_epjd_2017,sargsyan_lpl_2014}. For strong magnetic fields $B\gg B_0$ the probabilities of MI transitions tend to zero, contrary to the one of the IFFA transitions that  asymptotically tend to its maximum with increasing $B$-field (see Fig.~\ref{fig6}a). We have confirmed this statement using a magnetic field of 500~G and MI transitions of the $^{39}$K, $F_g=1\rightarrow F_e=3$ with $\sigma^+$ excitation. While IFFA transitions already have big amplitudes (see Fig.~\ref{fig5}), the MI transitions are not detectable. Note that for the confirmation of this statement one must to apply a magnetic field of 5--6~kG on $^{87}$Rb atoms.

\subsection{Discussion}

The peculiarities of the atomic transitions behaviour of  K in the HPB regime is different in the case D$_1$ and D$_2$ lines: (\textit{i}) as mention earlier there are two groups of eight transitions formed either by $\sigma^+$ and $\sigma^-$ circularly-polarised light for the D$_2$ line and each of these group contains one GT. Meanwhile, in the case of D$_1$ line there are two groups of only four transitions, one for each circularly-polarised light, and the GT are absent. (\textit{ii}) In the case of $\pi$-polarised laser radiation, the spectrum of K D$_2$ is composed by eight atomic transitions, including two IFFA transitions, meanwhile the one of the D$_1$ line counts two GT and two IFFA transitions.

\begin{figure}[ht]
\centering
\includegraphics[width=0.7\textwidth]{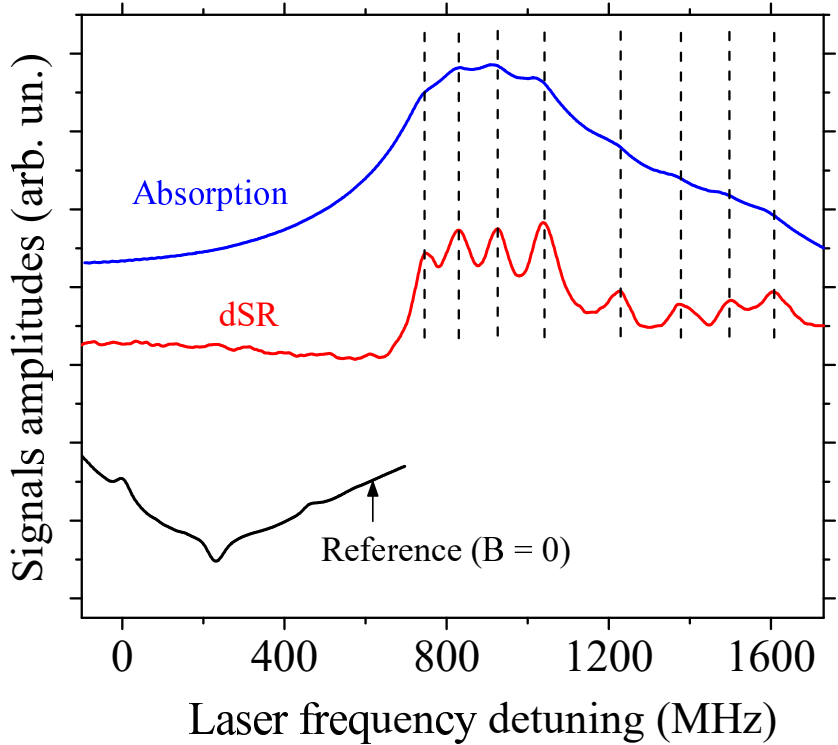}
\caption{$^{39}$K D$_2$ line for a $\sigma^+$- polarised radiation and a magnetic field $B=600$~G. The upper (blue) trace shows the absorption spectrum obtained with a NC of thickness $L=385$~nm; the middle (red) trace shows the corresponding dSR spectrum where eight atomic transitions are completely resolved. The lower curve is the SA spectrum that serves a frequency reference.}
\label{fig7}
\end{figure}

Investigation of the modification of the transition frequencies and probabilities for K D$_1$ line using absportion spectroscopy from a NC was reported in \cite{sargsyan_epl_2015}. The small number of atomic transitions (four for each circular polarisation) and the narrowing of the spectra due to the Dicke effect  at $L = \lambda/2 = 385$~nm \cite{sargsyan_jpb_2016} were the reasons that the resolution of the recorded spectra was sufficient enough. However, due to a larger number of the atomic transitions (eight) formed by circularly-polarised laser radiation, absorption spectra of K D$_2$ line in NC are strongly broadened. For this reason, we illustrate on Fig.~\ref{fig7} the advantage of the dSR technique over the conventional absorption one.  In this figure, the upper blue trace shows the absorption spectrum of K D$_2$ line obtained from a NC with $L=385$~nm and $\sigma^+$ laser excitation and a magnetic field set as $B\sim600$~G.  Although the eight absorption peaks are slightly resolved, they have big pedestals which overlap with one another, causing strong distortions in amplitudes. In order to get the correct ones, one needs to perform a not trivial fitting due to a special absorption profile inherent to the NC. Meanwhile, the middle curve shows the corresponding dSR spectrum, where eight atomic transitions are completely resolved. Thus, for K D$_2$ line selective reflection technique is strongly preferable.

As mentioned earlier, a remarkable property of K atoms is that the HPB regime occurs at very small magnetic field as compared to widely used Rb and Cs atoms, and magneto-optical studies using the HPB regime can be realized with a much smaller magnetic field. Particularly, in \cite{zlatkovic_lpl_2016}, a complete HPB regime was achieved for the $^{133}$Cs D$_2$ line for a magnetic field $B=27~$kG. Using an additional laser and optical pumping process of the ground-state sublevel a high polarisation of nuclear momentum was achieved. Similar results could be obtained with K atomic vapour for 10 times smaller magnetic field $B \sim 2.5$~kG, because $B_0 (^{39}$K$)/B_0(^{133}$Cs$) \sim 10$. 

Let us note that the energy level structure for the isotope $^{41}$K is very similar to that of $^{39}$K, while the hyperfine splitting for the ground and excited levels are smaller \cite{bendali_jpb_1981,tonoyan_nasra_2016}. Particularly, the hyperfine splitting of the ground $4S_{1/2}$ level is 254 MHz, which is 1.8 times smaller than the one of $^{39}$K. This means that the constant $B_0(^{41}$K)$~\sim90$~G and the HPB regime is achieved at smaller magnetic field. In addition, the structure of $^{41}$K spectrum  in strong longitudinal magnetic field is the same than $^{39}$K: two groups of eight transitions are recorded for circularly-polarised excitation, each of these groups containing one GT. In the case of $\pi$-polarised radiation, one can count eight transitions from which two are IFFA transitions. Transitions of $^{41}$K follow the same behaviour than the ones of $^{39}$K (see Fig.~\ref{fig3} and \ref{fig6}a), while equivalent probabilities are reached at a smaller magnetic field strength. 

\section{Conclusion}

We have demonstrated that, despite a large Doppler-width of atomic transitions of potassium, SR of laser radiation from a K atomic vapour confined in a NC with a 350 nm gap thickness allows to realize close to Doppler-free spectroscopy. Narrow linewidth and linearity of the signal response with respect to transition probabilities allow us to detect, in an external longitudinal magnetic field, separately two groups each of them composed by eight transitions and formed either by $\sigma^+$- or $\sigma^-$-polarised light. 

We have also showed that the dSR-method provided much better spectral resolution ($\sim80$~MHz) than that based on the absorption spectrum realized in NC with a thickness $L=\lambda/2$, since narrow linewidth of the transitions allows one to avoid overlapping of the nearly located  atomic transitions; here, the frequency separation between transitions is $\sim100$~MHz. 
	
The theoretical model describes the experiments very well. The experimental results along with calculated magnetic field dependence of the frequency shifts and the probabilities for 1--4 (1'--4') and 5--\textcircled{8} (\textcircled{5}'--8') transitions under $\sigma^+$ ($\sigma^-$) laser radiation, as well as the one for transitions 1--8 transitions observed in the case of $\pi$-polarised laser excitation and the detection of two IFFA transitions give the complete picture of potassium D$_2$ line atomic transitions behaviour in magnetic field. 
	              
	The implementation of this recently developed setup based on narrowband laser diodes, strong permanent magnets, and dSR method using NC allows one to study the behaviour of any individual atomic transition of $^{39}$K atoms as well as of $^{23}$Na, $^{85}$Rb, $^{87}$Rb, $^{133}$Cs D$_1$ and D$_2$ lines. Particularly, the NC-based dSR method could be a very convenient tool for the study of $^{23}$Na atomic vapour, since the atomic transitions are masked under a huge Doppler width of $\sim1.5$~GHz in conventional spectroscopic experiments.
	
	 It should be noted that a recently developped fabrication process of a glass NC \cite{whittaker_jpc_2015}, much simpler than the one using sapphire materials, will make the NC-based dSR-technique widely available for researchers.

\ack
The authors are grateful to A. Papoyan and G. Hakhumyan for valuable discussions.


\end{document}